\documentclass[10pt]{article}
\usepackage[OE]{express}
\usepackage{geometry, amssymb, amsbsy, amsmath, latexsym, dsfont, array, layout, graphicx, mathrsfs,color,bbm,endnotes}
\newcommand{\fig}[1]{Fig.~\ref{#1}}

\newcommand{\ket}[1]{\left|{#1}\right\rangle}

\newcommand{\eq}[1]{Eq.~(\ref{#1})}
\newcommand{\beq}{\begin{eqnarray}}
\newcommand{\eeq}{\end{eqnarray}}

\usepackage[normalem]{ulem}

\begin{document}
\title{Enhanced violations of Leggett-Garg inequalities in an experimental three-level system}

\author{Kunkun Wang,\authormark{1} Clive Emary,\authormark{2} Xiang Zhan,\authormark{1} Zhihao Bian,\authormark{1} Jian Li,\authormark{1} and Peng Xue, \authormark{1,3,*}}

\address{\authormark{1}Department of Physics, Southeast University, Nanjing 211189, China\\
\authormark{2}Joint Quantum Centre (JQC) Durham-Newcastle, School of Mathematics and Statistics, Newcastle University, Newcastle-upon-Tyne, NE1 7RU, United Kingdom\\
\authormark{3}State Key Laboratory of Precision Spectroscopy, East China Normal University, Shanghai 200062, China}

\email{\authormark{*}gnep.eux@gmail.com} 



\begin{abstract}
Leggett-Garg inequalities are tests of macroscopic realism that can be violated by quantum mechanics.  In this letter, we realise photonic Leggett-Garg tests on a three-level system and implement measurements that admit three distinct measurement outcomes, rather than the usual two. In this way we obtain violations of three- and four-time Leggett-Garg inequalities that are significantly in excess of those obtainable in standard Leggett-Garg tests.  We also report violations the quantum-witness equality up to the maximum permitted for a three-outcome measurement. Our results highlight differences between spatial and temporal correlations in quantum mechanics.
\end{abstract}



\section{Introduction}
In contrast to Bell inequalities which probe correlations between multiple spatially-separated systems \cite{Bell1964,Giustina2015,Shalm2015}, the Leggett-Garg inequalities (LGIs) test the temporal correlations of a single system \cite{LG1985,LeggettGarg1987,Clive2015}.
The LGIs are based on two macrorealistic assumptions that intuitively hold in the world of our everyday experience: (i) \textit{macroscopic realism per se} --- that a system exists at all times in a macroscopically-distinct state; and (ii) \textit{non-invasive measurability} --- that it is possible to measure a system without disturbing it.
Since both these assumptions fails under quantum mechanics, quantum systems can violate the LGIs. Hence the use of these inequalities as indicators of quantum coherence, in particular  in macroscopic systems \cite{Leggett2002}.

The LGIs concern the correlation functions $C_{ij} = \langle Q(t_i)Q(t_j)\rangle$ of dichotomic variable $Q(t)=\pm 1$ at measurement times $\left\{t_i\right\}$.  Typical three- and four-time LGIs can be written \cite{Clive2015}
\beq
  -3 \le K_3\le 1
  ;\quad
  K_3 \equiv C_{21} + C_{32} - C_{31} \label{EQ:K3}
  \\
  |K_4| \le 2
  ;\quad
  K_4 \equiv C_{21} + C_{32} + C_{34} - C_{41} \label{EQ:K4}
  ,
\eeq
as temporal analogues of the original Bell \cite{Bell1964} and CHSH \cite{Clauser1969} inequalities. These inequalities, in particular the $K_3$ inequality and its close relatives, have been tested and violated in many experiments, with most studies having been performed on two-level quantum systems, e.g.  \cite{PLaloy2010,Waldherr2011,Athalye2011,Xu2011,Goggin2011,Dressel2011,KneeSimmonsGauger2012,Groen2013,Katiyar2013,Asadian2014,Zhou2015,Knee2016,Formaggio2016a}.
In such systems, the maximum quantum-mechanical value of the Leggett-Garg (LG) correlators are  $K^\mathrm{TTB}_3 = 3/2$ and $K^\mathrm{TTB}_4 = 2\sqrt{2}$ in the three- and four-time case respectively. The derivation of these values is analogous \cite{Fritz2010} to that of the Tsirelson bounds \cite{Cirelson1003,Poh2015,MN16} of the corresponding inequalities for spatially separated observations and we will refer to these bounds as the {\em temporal Tsirelson bounds} (TTBs) of the LGIs
(this bound was also referred to as the L{\"u}ders bound~\cite{Budroni2014}).
It is known that $K_{n}^\mathrm{TTB}$ bound the LGIs for quantum systems of arbitrary size provided that the measurements are genuinely dichotomic, i.e. can be modelled with exactly two projection operators \cite{Budroni2013}. Recently, however, it was predicted that values of $K_3$ exceeding $K_3^\mathrm{TTB}$ are possible for $N$-level systems when the measurement apparatus provides more information than a single bit, and is thus modelled with $M>2$ orthogonal projectors \cite{Budroni2014}.
In particular, for a three-level system with measurements decomposed as three projectors (each nevertheless associated with a value of either $Q=+1$ or $Q=-1$) it was predicted that the maximum value of the LG correlator under quantum mechanics is $K_3^\text{max} =2.1547$. A similar substitution of multi-outcome measurements into the Bell and CHSH  inequalities leaves the (spatial) Tsirelson bounds unaltered \cite{Budroni2014}.

A small number of experiments have been performed on multi-level systems \cite{George2013,Robens2015,KatiyarBrodutchLu2016}, but no violations $K_n > K_n^\mathrm{TTB}$ have yet been reported. In \cite{Robens2015} the decisive measurement at $t_2$ was only a two-outcome one. In \cite{George2013} three-outcome measurements were considered but the focus there was on ``non-disturbing measurements'', rather than on maximising $K$. Recently, a three-level NMR system was studied \cite{KatiyarBrodutchLu2016} for which a theoretical maximum violation of $K_3 = 1.7566>K_3^\mathrm{TTB}$ was predicted.  However, no evidence of violations greater than the TTB were found. Violations exceeding $K_3^\mathrm{TTB}$ have recently been theoretically studied in multi-qubit systems \cite{Lambert2016}. As far as we are aware, the violation of the CHSH-like $K_4$ LGI in excess of the TTB has not been observed experimentally.

In this paper, we report on LG experiments with single photons that implement a three-level quantum system measured with three orthogonal projectors. We investigate both $K_3$ and $K_4$ inequalities and our main result is the observation of maximum values of the LG correlators $K_3=1.97\pm0.06$ and $K_4=2.96\pm0.05$, which clearly represent significant enhancements over the TTB.
We also consider a quantum-witness \cite{LiLambertChen2012} (or no-signalling-in-time \cite{Kofler2013}) test for our system, which is based on the same assumptions as the LGIs but is simpler and is in some ways preferable \cite{ClementeKofler2016}. In contrast to our results for the LGI, where the measured violations are still lower than the theoretical maximum, our maximum measured value for the quantum witness saturates the theoretical bound for a three-outcome test \cite{SchildEmary2015}, and presents a significant enhancement over the hitherto-observed value for a two-outcome case \cite{Robens2015}.

These values we obtain using \textit{ideal negative measurements} (INMs) \cite{LG1985}. As in~\cite{KneeSimmonsGauger2012,Robens2015,KatiyarBrodutchLu2016}, these allow us to acquire information about the system (here, the photon) without interacting with it directly, and thus take steps to address the ``clumsiness loophole'' \cite{Wilde2011}. Our interferometeric set-up makes the designation of our measurements as INM extremely clear-cut.


\section{Multiple-outcome LGI tests}

\begin{figure*}[tbh]
\centering
   \fbox{\includegraphics[width=\textwidth]{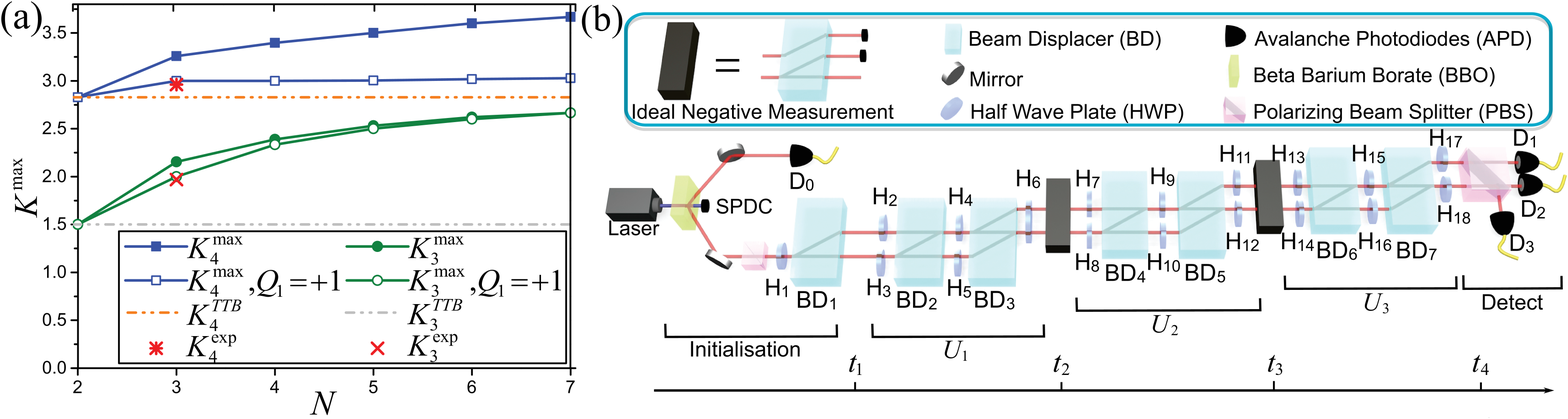}}
   \caption{
    \textbf{(a)} Theoretical maximum values of $K_3$ and $K_4$ for $M=N = 2\ldots7$. Results are shown both with and without restriction that the first measurement coincides with state preparation, $Q(t_1) = +1$.  Also indicated are the maximum experimental violations reported here.
    \textbf{(b)}
    Experimental set-up for the four-term CHSH-style LGI with the heralded single photons. The first PBS, HWP (H$_1$) and BD$_1$ are used to generate the initial qutrit state $\ket{C}$ at $t_1$.
    Sets of HWPs and BDs are used to realize the evolution operators $U_i$.
    Projective measurement of the final photon state at $t_3$ is realized via a PBS which maps the basis states of the qutrit into three spatial modes. Detecting heralded single photons means in practice registering coincidences between the trigger detector D$_0$ and each of the detectors for measurement D$_1$, D$_2$, and D$_3$.
    The ideal negative measurement at times $t_2$ and $t_3$ is realized by blocking channels, two at a time, such that detection at D$_\mathrm{1-3}$ implies that path taken was the non-blocked channel.
   }
\label{FIG:Experimental_setup}
\end{figure*}

We consider a system in which we measure a variable $m$ with $M$ distinct outcomes. We connect with the standard LGI framework by introducing the mapping of a measurement of $m_i$ at times $t_i$ onto the value $Q_i=Q(t_i)=q(m_i,t_i)\in\{-1,+1\}$. The correlation functions are thus constructed $
  \langle Q(t_i)Q(t_j)\rangle=\sum_{m_i,m_j}q(m_i,t_i)q(m_j,t_j)P_{ij}(m_i,m_j)
$, where $P_{ij}(m_i,m_j)$ is the joint probability to obtain $m_i$ and $m_j$ at times $t_i$ and $t_j$.
This way of constructing $Q$ and its correlation functiosn clearly leaves the classical upper bounds in \eq{EQ:K3} and \eq{EQ:K4} unaffected.
The maximum quantum-mechanical values for the $K_3$ and $K_4$ correlators are shown in  \fig{FIG:Experimental_setup}(a) as a function of the number of levels $N$ of the quantum-mechanical system with number of outcomes $M=N$. These values were obtained numerically as described in \cite{Budroni2013}.
For $M>2$, violations significantly higher than the TTB are clearly possible.
In the following, we will make use of a common experimental simplification of the LGIs \cite{Goggin2011,Robens2015,Lambert2016} and assume the coincidence of state preparation with measurement at $t_1$, and simply define $Q(t_1)=1$.  This can reduce the maximum quantum violations, as shown in \fig{FIG:Experimental_setup}(a).

%


\section{Experimental Realization}
\subsection{Experimental Set-Up}
We consider the smallest system that will admit a three-outcome measurement, a qutrit with states $\ket{n};~n=A,B,C$ ($M=N=3$), which we realise with single photons, Fig.~\ref{FIG:Experimental_setup}(b). The basis states $|A\rangle$, $|B\rangle$, and $|C\rangle$ are encoded respectively by the horizontal polarization of the heralded single photon in the upper mode, the horizontal polarization of the photon in the lower mode, and the vertical polarization of the photon in the lower mode.  Heralded single photons are generated via a type-I spontaneous parametric down-conversion (SPDC). The polarization-degenerate photon pairs are produced in SPDC using a $\beta$-barium-borate (BBO) nonlinear crystal pumped by a diode laser. With the detection of a trigger photon the signal photon is heralded for evolution and measurement \cite{Xue2015,BQZ+15,ZZL+16,XZB+17}.  The pump is filtered out with the help of an interference filter which restricts the photon bandwidth to $3$nm.

Initial qutrit states are prepared by first passing the heralded single photons through a polarizing beam splitter (PBS), and a half-wave plate (HWP, H$_1$) before being split by a birefringent calcite beam displacer (BD) into two parallel spatial modes, upper and lower, with vertically-polarized photons directly transmitted through the BD in the lower mode, and with horizontal photons undergoing a $3$mm lateral displacement into a upper mode. In the current set-up we set PBS and H$_1$ to give vertically-polarized photons, which then remain in the lower mode through BD$_1$, thus initialising the qutrit in the state $\ket{C}$.

Time evolution from $t_i$ to time $t_{i+1}$ in our $K_4$ experiment is given by unitary operators $U_{i}$ ($i = \{1,3\}$),
\begin{align}
  U_i&=\left(\begin{array}{ccc}
      \cos\theta_i & 0 & \sin\theta_i \\
				      \sin\theta_i\sin\phi_i & \cos\phi_i & -\cos\theta_i\sin\phi_i \\
				      -\sin\theta_i\cos\phi_i & \sin\phi_i & \cos\theta_i\cos\phi_i \\
  \end{array}\right),
\end{align}
in the basis ($\ket{A}=(1,0,0)^\text{T},\ket{B}=(0,1,0)^\text{T},\ket{C}=(0,0,1)^\text{T}$). The middle unitary between $t_2$ and $t_3$ is then set to $U_2=U^\dagger_3 U^\dagger_1$. For the $K_3$ test the time evolution operator $\tilde{U}_1$ between $t_1$ and $t_2$ is taken the form shown in Eq.~(3) and the evolution operator between $t_2$ and $t_3$ is set to $\tilde{U}_2=\tilde{U}^\dagger_1$. These are realised by a sequence of HWPs and BDs with $\theta_i$ and $\phi_i$ adjustable parameters~\cite{Reck1994,ZQL+15,Wang2016,ZCL+17,XWZ+17}.

We identify our measurements as projections onto the three basis states with outcomes $m\in\left\{A,B,C\right\}$ and thus the probability $P(m,t)$ of obtaining outcome $m$ is the same as the probability $P(n,t)$ of detecting the system in state $n$.
Projective measurement of the final photon state is performed with a PBS that maps the quitrit basis states of qutrit into three spatial modes and to accomplish the projective measurement. The photons are then detected by single-photon avalanche photodiodes (APDs), in coincidence with the trigger photons. The probability of the photons being measured in $\ket{n}$ $(n=A,B,C)$ is obtained by normalizing photon counts in the certain spatial mode to total photon counts.  The count rates are corrected for differences in detector efficiencies and losses before the detectors. We assume that the lost photons would have behaved the same as the registered ones (fair sampling). Experimentally this trigger-signal photon pair is registered by a coincidence count at APD with $7$ns time window
(total coincidence counts were approximately $28,000$ over a collection time of $14$s in $K_4$ test).

INM of the qutrit state at earlier times is realised by placing blocking elements into the optical paths~\cite{Emary2012a}. With, for example, the channels $A$ and $B$ blocked at $t_2$, we obtain the probabilities $P_{42}(n_4, n_2 = C)$ without the measurement apparatus having interacting with the photon. In our experiment, this blocking is realized by a BD followed by an iris. The BD is used to map the basis states of qutrit to three spatial modes and the iris is used to block photons in two of the three spatial modes and let the photons in the rest one pass through. By changing the position of the iris, we can block any two of the channels and let the photons in the remaining one pass through for the next evolution. By using different sequences of blocking and unblocking as well as final detection at $t_4$, all the necessary correlation functions can be constructed.

\subsection{CHSH-Type Inequality}

We first consider $K_4$ with $Q(t_1)\equiv 1$ and $q(A,t_i)=q(B,t_i)=1$ and $q(C,t_i)=-1$ for $i=2,3,4$.
We choose the middle unitary $U_2$ according to $U_2=U^\dagger_3U^\dagger_1$, such that $U_3U_2U_1=\mathbbm{1}$, where $\mathbbm{1}$ is a $3\times3$ identity matrix and set $\theta_1=\pi/2$ and $\phi_1=0$.
In this case, the theoretical value of the LGI $K_4$ correlator as a function of $\phi_3$ and $\theta_3$ is $K_4=\left[9+2\cos(2\theta_3)\cos(4\phi_3)-2\cos(4\phi_3)-2\cos(2\theta_3)\right]/4$.
This takes a maximum value of $K_4 = 3$ for $\theta_3=\pi/2$ and $\phi_3=\pi/4,3\pi/4$ which is the maximum value achievable under the condition that the $t_1$-measurement coincides with initialisation.

\begin{figure}
\centering
   \fbox{\includegraphics[width=0.65\columnwidth,clip=true]{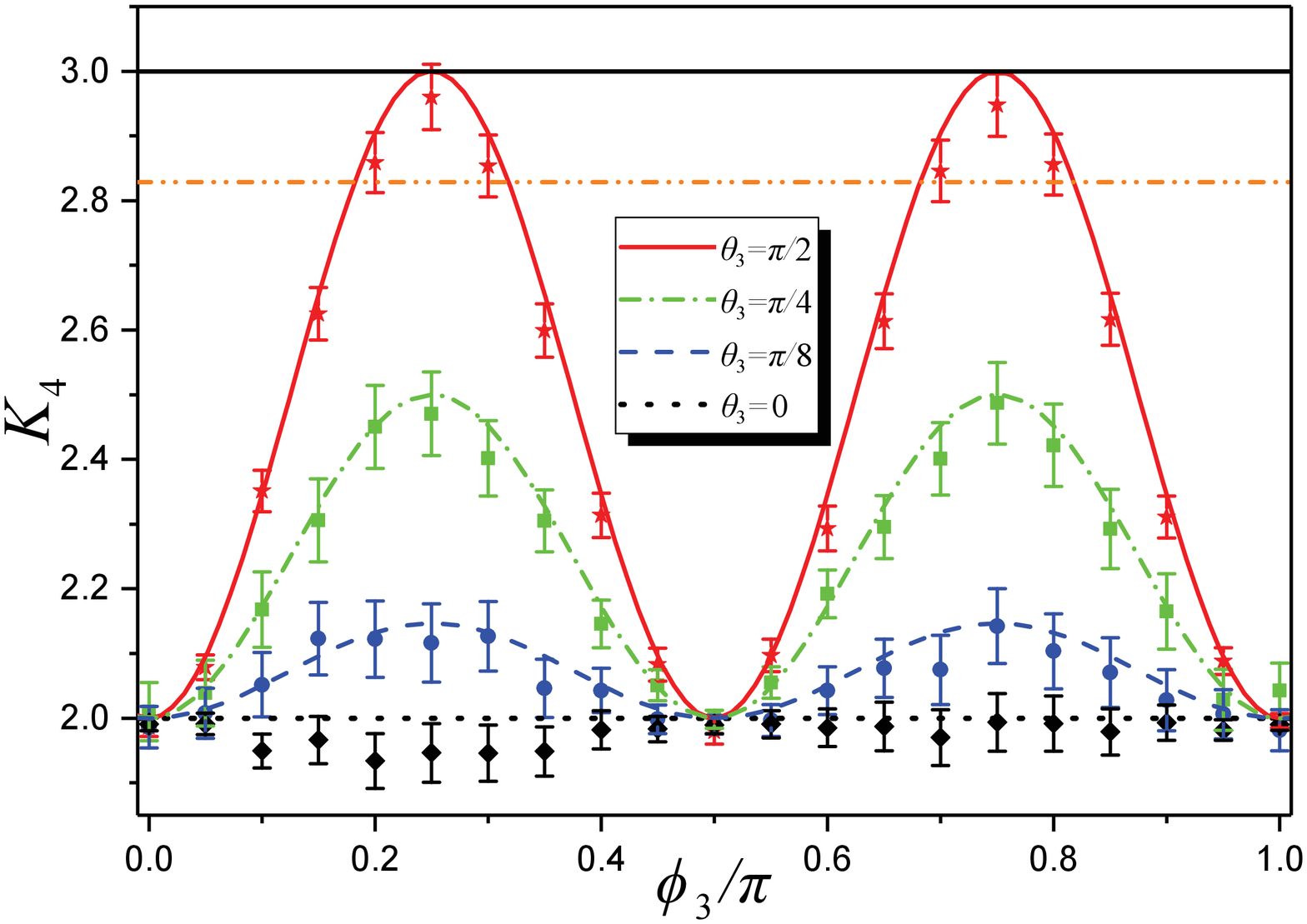}}
   \caption{
     Experimentally-determined values of the four-time CHSH-style Leggett-Garg correlator $K_4$ for our three-level, three-outcome set-up with time evolution described by parameters $\phi_3$ and $\theta_3$. Theoretical predictions are represented by different lines
     and the experimental results by symbols.
     Error bars indicate the statistical uncertainty based on the assumptions of Poissonian statistics.
     The maximum measured value of the Leggett-Garg correlator occurs at $\phi_3=\pi/4$ ($\phi_3=3\pi/4$) and has value $K_4 = 2.96\pm0.05$ ($2.95\pm0.05$), as compared with the theoretical maximum value of $3$ (represented by the black solid line).
     This represents a significant enhancement over the temporal Tsirelson bound of $2\sqrt{2}$ (represented by the orange dash-dot-dot line).
     \label{FIG:K4}
   }
\end{figure}

Figure~\ref{FIG:K4} shows our experimentally determined value of $K_4$ as a function of $\phi_3$ for several values of $\theta_3$.
Agreement with theory is close, and the maximum violations obtained are $K_4=2.96\pm0.05$ for $\phi_3=\pi/4$ and $K_4=2.95\pm0.05$ for $\phi_3=3\pi/4$.
These values thus show clear experimental evidence of the super-Tsirelson-bound violations of the four-term CHSH-style LGI. Error bars indicate the statistical uncertainty, based on the assumption of Poissonian statistics. Compared with the theory, the measured value of $C_{21}$ is close to its theoretical prediction since the joint probability of $P_{21}(n_1=C,n_2)=\sum_{n_4}P_{42}(n_4,n_2)$ is less affected by the imperfection of cascaded interferometers after the INM at $t_2$. Furthermore, there is no cascaded interferometer in the measurement of $C_{34}$ and $C_{32}$, such effects are reduced and the measured values of $C_{34}$ and $C_{32}$ are close to their theoretical predictions too. The main deviation from theory arises in the measurement of $C_{41}$ where, by construction, the final state should be the same as the initial state such that $C_{41}=-1$. However, due to imperfection in the cascaded interferometers in this setup, $P_4(n_4=C)$ is smaller than $1$ and we obtain $C_{41}=-0.975 \pm0.002$ at $\{\theta_3,\phi_3\}=\{\pi/2,\pi/4\}$.
There are about $20$ pieces of wave plate used in the setup and each of them has an angle error of approximately $0.1^\circ$.  In a cascaded setup, these errors accumulate. Monte Carlo simulation of this scenario show that these angular errors are sufficient to explain the deviation of experimental results from their theoretical predictions~\cite{WEX+17}.


\subsection{Three Term LGI and Quantum Witness}

\begin{figure}
\centering
  \fbox{\includegraphics[width=.65\columnwidth,clip=true]{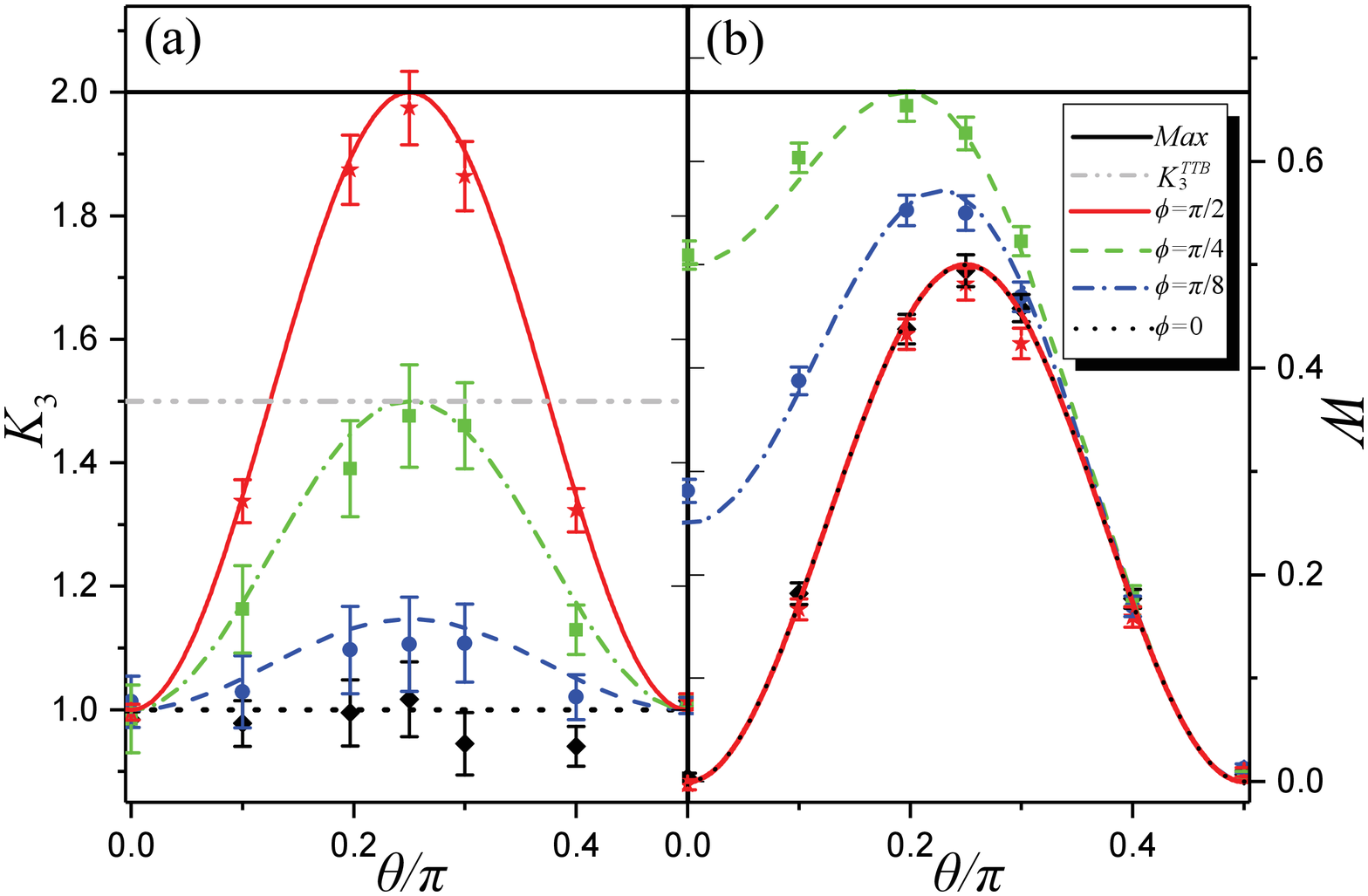}}
   \caption{
     \textbf{(a)}  As \fig{FIG:K4} but for the three-time Leggett-Garg correlator $K_3$.
     The maximum measured value of the Leggett-Garg correlator occurs at $\{\phi,\theta\}=\{\pi/2, \pi/4\}$ and has value $K_3 = 1.97\pm0.06$, as compared with the theoretical maximum value of $2$.
     \textbf{(b)}  The experimentally-determined values of the quantum witness $W$
     for our three-level, three-outcome setup. The maximum value of the quantum witness is $W = 0.65\pm0.02$, which occurs for evolution parameters $\{\phi,\theta\}= \{\pi/4,\arccos\sqrt{2/3}\}$. This saturates the theoretical maximum of $W=2/3$ for a three-level system. The maximum value of the corresponding LGI, \eq{EQ:witness}, is $K_{3W}=1.65\pm0.02$. Both the maximum violation of $K_3$ and $K_{3W}$ exceed the TTB $K^{TTB}_3=3/2$.
     \label{FIG:K3nW}
   }
\end{figure}

The experimental geometry for the $K_3$ is essentially the same as above with one time-evolution step removed.
In this case, and following~\cite{SchildEmary2015}, we take the second time evolution to be $\tilde{U}_2=\tilde{U}_1^\dagger$ such that $\tilde{U}_2\tilde{U}_1=\mathbbm{1}$.  This gives the correlation functions
$C_{31}=-1$,
$C_{21} = \sin^2 \theta-\cos^2 \theta \cos 2\phi$, and
$
  C_{32} = \cos 2\theta \left[\sin^2 \theta + \cos^2 \theta (\cos^4 \phi-\sin^4 \phi)\right]
  \nonumber
$.
The maximum value of the corresponding LGI is $K_3=2$, which occurs for the evolution parameters $\theta=\pi/4,3\pi/4$ and $\phi=\pi/2$.
Figure~\ref{FIG:K3nW}(a) shows the observed values of $K_3$ as functions of $\theta$ and $\phi$.  Maximum violation is found for $\phi = \pi/2$ with a value $ K_3 = 1.97\pm0.06$ at $\theta = \pi/4$ and $K_3 = 1.95\pm0.06$ at $\theta = 3\pi/4$ (not shown), in close agreement with the theoretical prediction.

Whilst neither of the above LGI measurements saturate the maximum theoretical value of the LGI with $N=M=3$ shown in \fig{FIG:Experimental_setup}(a), we are able to saturate the $M=3$ bound for the quantum witness
\begin{equation}
  W\equiv P_3(m_3=C)-\sum_{m_2}P_{32}(m_3=C,n_2)
  ,
\end{equation}
based on registering the outcome $m_3=C$. Under macrorealism and non-invasive measurability, we have the equality $W=0$.
This witness can be constructed from the same probabilities as used in the $K_3$ test and these results are shown in \fig{FIG:K3nW}(b). Theory \cite{SchildEmary2015} predicts a maximum value of
this witness should occur for the parameters $\phi=\pi/4$ and $\theta = \arccos\sqrt{2/3}$ and $\theta = \pi - \arccos\sqrt{2/3}$. At these points we observe the values $W = 0.65\pm0.02$ and $W = 0.64\pm0.02$. This agrees well with the theoretical value of $W = 1-M^{-1}=2/3$, which is the maximum possible value for a three-projector measurement (and thus, also for a three-level system).

We can relate this quantum witness directly to a LGI if we choose measurement value assignments $q(m_2,t_2)=1$ at $t_2$ (a blind measurement) \cite{Robens2015}, and $q(m_3,t_3)=\delta_{m_3,C}$. In this case, the LGI of \eq{EQ:K3} reduces to
\beq
  K_{3W} = 1+W\leq1
  \label{EQ:witness}
  .
\eeq
Thus, the value by which the witness exceeds zero is the extent to which the corresponding LGI is violated.  The maximum violation of this LGI with this measurement assignment is thus $K_{3W}=1.65\pm0.02$.

There is connection between enhanced violations of the LGI, quantum witness equality and dimension witnesses~\cite{Budroni2014,SchildEmary2015,H12,A12}. However, from the view of experiment, the dimension of the system being measured is usually known before we can design an experimental setup. Therefore the dimensionless witness test is not considered here.

\section{Discussion}

We have demonstrated experimental violations of LGIs in a three-level system and obtained values of the LG correlators $K_3$ and $K_4$ greatly in excess of the TTBs, familiar from studies of two-level systems. We have also demonstrated a similar excess for the quantum-witness equality.
These enhancements arise because the decisive $t_2$- and $t_3$-measurements here admit three distinct measurement outcomes, rather than the usual two. Under this measurement, the collapse of the wave function is greater than with two projectors and the resultant additional information gain enables the enhanced violation. In particular, in the case of the witness, the post-measurement state of the qutrit is the maximally mixed state $\rho = \frac{1}{3}\times \mathbbm{1}$. The corresponding violation is therefore up to the theoretical maximum for three-outcome measurements. These results provide an experimental demonstration of the difference between spatial and temporal correlations in quantum mechanics, since the (spatial) Tsirelson bound in the Bell and CHSH inequalities remains fixed at $3/2$ and $2\sqrt{2}$, irrespective of the number of projectors.

In the future, it will be interesting to look at temporal analogues of different (spatial) Bell inequalities, in particular ones with multi-outcome measurements \cite{Acin2006,Hirsch2016}. A further interesting area is the investigation of how the maximum quantum violations of the LGIs scale with increasing system size, where theoretical results suggest that the algebraic bound is obtainable in the asymptotic limit \cite{Budroni2014}.  We note that our measurements are within the standard quantum-mechanical framework and thus demonstrate that post-quantum effects are not needed to obtain enhanced LGI violations \cite{Dakic2013}.

Classical invasive measurements can give violations of the LGI, all the way up the algebraic bound \cite{Montina2012}. It is therefore important to ensure the non-invasivity of the measurements in any LGI test.  Whilst no known scheme can completely rule out such invasivity
(results such as in~\cite{Wilde2011} and \cite{Knee2016} reduce the ``size'' the clumsiness loophole, but do not close it altogether),
we have used INMs here which rule out the direct influence of the measurements on the system itself. Nevertheless, our measurements of the quantum witness indicate that the correlations here are of the ``signalling'' type \cite{Kofler2013,Halliwell2016}, which then points to an interesting comparison between our work, where we have  both signalling and $K_n>K_n^\mathrm{TTB}$, and in \cite{George2013} where no-signalling was obeyed but the LGI violation was restricted to $K<K_n^\mathrm{TTB}$.

Finally, we note that these results demonstrate the versatility of beam-displacer interferometer networks, as well as such techniques as full control of both polarization and spatial modes of single photons, for use in fundamental tests of quantum mechanics beyond the Bell inequalities. From this perspective, the LGIs can be seen as a useful benchmark for quantum control experiments.

\section*{Funding}
National Natural Science Foundation of China (Nos.~11474049 and 11674056); Natural Science Foundation of Jiangsu Province (BK20160024); Open Fund from State Key Laboratory of Precision Spectroscopy of East China Normal University; Scientific Research Foundation of the Graduate School of Southeast University.

\section*{Acknowledgments}

We would like to thank Neill Lambert, George Knee and Jonathan Halliwell for helpful discussions.

\end{document}